\documentclass[aps,twocolumn]{revtex4}
\usepackage [dvips]{graphicx}
\begin{document}

\title{Nuclear Superfluidity and Cooling Time of Neutron-Star Crust}

\author{
C. Monrozeau$^{a}$, J. Margueron$^{a}$, 
N. Sandulescu$^{a,b}$\footnote{corresponding author (email: sandules@ipno.in2p3.fr)}
}

\affiliation {
$^{a}$ Institut de Physique Nucl\'eaire, Universit\'e Paris Sud F-91406 Orsay CEDEX, 
France \\
$^{b}$ Institute of Physics and Nuclear Engineering, 76900, Bucharest, Romania}

\begin{abstract}
We analyse the effect of neutron superfluidity on the cooling time
of inner crust matter in neutron stars, in the case of a rapid cooling
 of the core. The specific heat of the inner crust, 
which determines the thermal response of the crust, is calculated in the
 framework of HFB approach at finite temperature. The calculations are 
performed with two paring forces chosen to simulate the pairing properties
 of uniform neutron matter corresponding respectively to Gogny-BCS approximation 
and to many-body techniques including polarisation effects.
Using a simple model for the heat transport across the inner crust, it
is shown that the two pairing forces give very different values for the
cooling time. 

\end{abstract}

\maketitle

\section{Introduction}

A newly-formed neutron star cools within minutes from a temperature of the order
of 30 MeV to less than 1 MeV via neutrino emission. After this stage, the thermal evolution
of the neutron stars can be strongly influenced by the onset of nuclear superfluidity
\cite{yakovlev,lattimer}.
This is especially the case for rapid cooling models. In these models, due to direct Urca
or other exotic processes, the core cools down so rapidly that a temperature inversion develops
between the core and the crust. The crust acts as an insulating blanket which keeps the surface
relatively warm until the cooling wave reaches the surface. When this happens, the surface
temperature drops precipitously to the temperature of the core. One of the relevant quantity
in this cooling scenario is the cooling time, i.e., the time necessary for the cooling 
wave to arrive from the cold core to the star's surface. The cooling time is primarely
determined by the
thermal response of the inner crust, formed by nuclear clusters immersed in a sea of unbound
neutrons and ultrarelativistic electrons \cite{pethick}.

In the rapid cooling models, both the core cooling and the time needed for the core-crust
thermalisation depend critically on nuclear superfluidity. Thus, on one hand, the onset of
superfluidity in the core matter suppresses the neutrino cooling since the total energy of
particles involved in the neutrino production must exceed the pairing gap. On the other hand,
the superfluidity of inner crust matter is shortening significantly the cooling time. This
happens due to the suppression of the heat capacity of the inner crust matter by the 
energy gap in the excitation spectrum of the superfluid neutron gas.

One of the first estimation of the cooling time was given by Brown et al \cite{brown}, who 
considered the
possibility of a rapid cooling induced by the strangeness condensation. 
They calculated the
heat diffusion time through the crust with a simple formula, i.e., $t_{diff} =\frac{R_c^2 
C_v}{\kappa}$, where $R_c$ is the thickness of the crust, while $C_v$
and $\kappa$ are the specific heat and the thermal conductivity of the crust matter.
The estimated cooling time was of the order of a few tens of years. Later on, using a direct
Urca process as cooling mechanism, more realistic calculations of thermal evolution of neutron
stars were performed \cite{lattimer}.
The numerical simulations showed that the cooling time does not depend on the details of
the rapid cooling mechanism but rather on the structure of the neutron star. Besides, it 
was also shown that the cooling time can be strongly reduced (by about a factor of three)
if the neutron gas in the inner crust is in a superfluid phase.

In the calculations mentioned above, the effects of nuclear clusters on the superfluid and
thermal properties of the neutron gas were disregarded. Since then, a few quantum calculations
of the inner crust matter superfluidity, including the effects of the nuclear clusters, have
been done \cite{barranco,sandulescu1,sandulescu2}. Thus, using the Hartree-Fock-Bogoliubov 
(HFB) approach it was found that the presence  of the nuclear clusters can modify significantly
the heat capacity of the neutron gas. How the nuclear clusters could 
affect the cooling time of the
 inner crust was investigated by Pizzochero et al \cite{pizzochero}.
 Using a cooling model similar to the one employed by Brown et al. \cite{brown},
it was concluded that the presence of the clusters, primarily in the outermost layers of 
the inner crust, could change the cooling time by amounts comparable with the cooling 
time itself. It was also found that the effect of the clusters on the cooling time
depends rather strongly on the temperature and the pairing force used in calculating
the specific heat of the inner crust matter.

The impact which the pairing force could have on the superfluid
properties and the specific heat of the inner cust matter was recently analysed in the framework
of HFB approach at finite temperature (FT-HFB) \cite{sandulescu2}. Thus, it was shown that if the
pairing force used in the FT-HFB equations is adjusted to describe two different scenarios
for the neutron matter superfluidity, i.e., one corresponding to BCS calculations with the
Gogny force and the other to Gorkov type calculations which  take into account self-energy
and screening effects \cite{shen}, the results for the specific heat of the inner crust matter
can change by several orders of magnitude. The scope of the present paper is to show what are 
the consequences of these changes in the specific heat upon the cooling 
time of inner crust matter.
In the first part of the paper we shall extend the calculations of Ref. \cite{sandulescu2} to the
low-density region of the inner crust, which was not treated before in the HFB approach, and analyse
how the specific heat and the thermal diffusivity behave across the inner crust. Then, using the 
model of Refs. \cite{brown,pizzochero} for the heat transport, we shall discuss how the cooling
time of the inner crust depends on neutron matter superfluidity.

\section{Thermal properties of inner crust matter in the HFB approach}

The thermal response of the inner crust matter depends on thermal diffusivity,
defined as the ratio of the thermal conductivity to the heat capacity. 
The heat capacity of the inner crust has contributions from the electrons, the
neutrons and the lattice. The heat capacity of the electrons, considered as a
uniform and relativistic gas, has the standard form \cite{landau} while the
contribution of the lattice to the specific heat is usually neglected. 

In the normal phase, the specific heat of the neutrons exceed the specific 
heat of the electrons by about two orders of magnitude (see Figure 1 below). 
However, the onset of 
the neutron superfluidity reduces drastically the neutron specific heat, which can 
become smaller than the electron specific heat in some regions of the inner crust.
How the neutron specific heat is affected by the superfluidity as well as by 
the temperature and the presence of nuclear clusters was already studied in
Ref. \cite{sandulescu2}, but only for a few density regions of the inner crust.
Here we extend this study to all relevant densities of the inner crust, 
starting
from the neutron drip density  up to about half the nuclear saturation density.
This region of the inner crust is supposed to give the largest contribution
to the cooling time of the crust \cite{pizzochero}.

\begin{table}[htb]
\centering
\begin{tabular}{c c c c c c}
\hbox{$N_{zone}$} & N &Z &$R_{WS}$&$\rho$ & $x_{i}$\\
 & & &$[fm]$& $[g.cm^{-3}]$& $[m]$ \\
\hline
\hline
10 & 140 &40  &54 &$4.7\times 10^{11}$&  12 \\
 9 & 160 &40  &49 &$6.7\times 10^{11}$&  12 \\
 8 & 210 &40  &46 &$1.0\times 10^{12}$&  15 \\
 7 & 280 &40  &44 &$1.5\times 10^{12}$&  21 \\
 6 & 460 &40  &42 &$2.7\times 10^{12}$&  40 \\
 5 & 900 &50  &39 &$6.2\times 10^{12}$&  45 \\
 4 &1050 &50  &36 &$9.7\times 10^{12}$&  43 \\
 3 &1300 &50  &33 &$1.5\times 10^{13}$&  87 \\
 2 &1750 &50  &28 &$3.3\times 10^{13}$& 156 \\
 1 &1460 &40  &20 &$7.8\times 10^{13}$& 187 \\
\end{tabular}
\caption{The Wigner-Seitz cells considered in the paper. The structure
of the cells, i.e., the baryonic densities ($\rho$), the number of neutrons (N),
the number of protons (Z) and the cell radii ($R_{WS}$) correspond to Ref. \cite{negele}.
$x_i$ are the thickness of the layers employed in Eq.(10).}
\label{tab3}
\end{table}

In microscopic calculations the inner crust matter is divided in independent
cells treated in Wigner-Seitz approximation \cite{negele}. Up to baryonic densities of 
the order of half the nuclear saturation density, considered in this paper, each cell
is supposed to contain in its center a spherical neutron-rich nucleus surrounded
by unbound neutrons and immersed in a relativistic electron gas uniformly 
distributed inside the cell. The proton- to- neutron ratio and the dimension
of the cell at a given baryonic  density are determined from the beta 
equilibrium conditions. In the present study we use the cell structure 
determined in Ref. \cite{negele} by HF type calculations. The properties of the 
cells considered in this paper are displayed in Table I. Compared to Ref. \cite{negele},
here we have not included the cell with Z=32, which most probably belongs to the deformed
pasta phase. For the cells listed in Table I we shall determine the specific heat by using
the quasiparticle spectrum generated by the FT-HFB approach presented below.

\subsection{The HFB approach at finite temperature}

The FT-HFB approach for the inner crust matter was presented 
in details in Ref. \cite{sandulescu2}. For the sake of completeness, 
here we recall the main steps.

Assuming spherical symmetry for the Wigner-Seitz cell, the radial FT-HFB 
equations have the form:
\begin{equation}
\begin{array}{c}
\left( \begin{array}{cc}
h_T(r) - \lambda & \Delta_T (r) \\
\Delta_T (r) & -h_T(r) + \lambda 
\end{array} \right)
\left( \begin{array}{c} U_i (r) \\
 V_i (r) \end{array} \right) = E_i
\left( \begin{array}{c} U_i (r) \\
 V_i (r) \end{array} \right) ~,
\end{array}
\label{1}
\end{equation}
where $E_i$ is the quasiparticle energy, $\lambda$ is the 
chemical potential, $h_T(r)$ is the thermal averaged mean field 
hamiltonian and $\Delta_T (r)$ is the thermal averaged
pairing field. The latter depends on the average pairing
density $\kappa_T$ given by:
\begin{equation}
\kappa_T(r)=\frac{1}{4\pi} \sum_{i} (2j_i+1) U_i^* (r) V_i (r)
(1 - 2f_i )~, \nonumber
\end{equation}
where $f_i = [1 + exp ( E_i/k_B T)]^{-1}$ is the Fermi distribution, $k_B$ is the
Boltzmann constant and T is the temperature.
 In a self-consistent calculation based on a 
Skyrme-type force, as used here, $h_T(r)$ depends on the thermal
averaged particle density
\begin{equation}
\rho_T(r) =\frac{1}{4\pi} \sum_{i} (2j_i+1) [ V_i^* (r) 
V_i (r) (1 - f_i ) \\
+ U_i^* (r) U_i (r) f_i ]~, 
\nonumber
\end{equation}
as well as on thermal averaged kinetic energy density and spin density.
The expressions of the last two densities are given in Ref. \cite{sandulescu2}.

In the calculations presented here the mean field hamiltonian is calculated with a Skyrme
type force while for the thermal averaged pairing field we use a  density  dependent 
contact force  of the following form \cite{bertsch}:
\begin{equation}
V (\mathbf{r}-\mathbf{r^\prime}) = V_0 [1 -\eta 
(\frac{\rho(r)}{\rho_0})^{\alpha}] 
\delta(\mathbf{r}-\mathbf{r^\prime}) 
\equiv V_{eff}(\rho(r)) \delta(\mathbf{r}-\mathbf{r^\prime}) ,
\end{equation}
where $\rho(r)$ is the baryonic density and $\rho_0=0.16$ fm$^{-3}$.
 With this force the thermal averaged pairing field is local
and given by:
\begin{equation}
\Delta_T(r) = \frac{ V_{eff}(\rho(r))}{2} \kappa_T (r) ,
\end{equation}
where $\kappa_T(r)$ is the thermal averaged pairing density.

To generate in the outer region of the Wigner-Seitz cell a constant density
corresponding to the neutron gas, the FT-HFB equations are 
solved by imposing Dirichlet-Neumann boundary conditions
at the edge of the cell \cite{negele}, i.e., all wave
functions of even parity vanish and the derivatives of 
odd-parity wave functions vanish. Apart from that, the 
self-consistent solutions of the HF-HFB equations are found in the
same manner as for finite nuclei. 

The calculation scheme outlined above is employed to study how the specific
heat of the neutrons is behaving in various regions of the inner crust.
In order to do that, one has to choose the two-body interactions in the FT-HFB
calculations. These interactions should provide a reasonable description of
both the nuclear clusters and the neutron gas, which are the baryonic 
components of the inner crust matter. For the calculation of the mean
field we shall use the Skyrme force SLy4 \cite{sly4}, which was fixed to describe
properly the mean field properties of neutron-rich nuclei and infinite
neutron matter. 

The choice of the pairing force is more problematic since
at present it is not yet clear what is the strength of pairing correlations
in neutron matter. Thus, on one hand, the BCS calculations
with bare forces give a maximum gap in neutron matter of about 3 MeV \cite{lombardo}. 
A maximum gap of about 3 MeV one gets also with the Gogny force \cite{gogny}, which is 
commonly used to describe
the pairing properties in finite nuclei. On the other hand, if one goes beyond
the BCS approximation and takes into account the in-medium effects, the maximum gap
is suppressed. The suppression depends on the many-body approximations used in the
calculations \cite{lombardo}. In order to analyse how the uncertainty on the pairing
gap in neutron matter could reflect upon the thermal response of the inner crust,
we shall do calculations with two zero range pairing interactions which simulate
the pairing gap in nuclear matter obtained either with the Gogny force, or with models which
take into account the in-medium effects. For the latter we consider a maximum gap of 1 MeV, as 
indicated by  recent calculations \cite{shen}. In Ref. \cite{sandulescu2} the requirements 
mentioned above were approximatively satisfied by using two zero range pairing forces (Eq.4)
having the same parameters for the density dependent term, i.e.,  $\eta$=0.7, $\alpha$=0.45, and 
two different strengths, i.e., $V_0$= \{-430.0,-330.0\} MeV fm$^{-3}$. These values of the strengths
were obtained by solving the FT-HFB equations with a cut-off energy equal to 60 MeV.
Since with a 60 MeV cut-off we have numerical problems in solving the FT-HFB equations for large Wigner-Seitz
cells, here we shall keep this cut-off and the corresponding strengths only for the first two cells while
for the other cells we shall take a smaller cut-off, equal to 20 MeV. This cut-off is introduced smoothly, 
i.e., by an exponential factor  $e^{-E^2_i/100}$ acting for quasiparticle energies $E_i>20$ MeV. With this
smooth energy cut-off we shall use the strengths values  $V_0$=\{-570.0,-430.0\} MeV fm$^{-3}$. The pairing
force corresponding to the first (second) value of the strength will be called below the strong (weak) 
pairing force.

\subsection{ Specific heat}

The quasiparticle spectrum determined by solving the FT-HFB
equations is  used to calculate the specific heat of the neutrons
inside the Wigner-Seitz cell, i.e.,
\begin{equation}
C_{V}=\frac{T}{V}\frac{\partial S}{\partial T} , 
\end{equation}
where V is the volume of the Wigner-Seitz cell and $S$ is the entropy:
\begin{equation}
S=-k_B \sum_{i} (2j_i+1) (f_{i} \ln f_{i}+(1-f_{i})\ln (1-f_{i})).
\end{equation}

 The results obtained for the cells listed in Table I are shown in Figure 1.
 In the same figure is also shown the specific heat of the electrons, 
 given by \cite{landau}:
 \begin{equation}
 C^{(e)}_V=\frac{k_B(3\pi )^{2/3}}{3\hbar c} \left( \frac{Z}{V} \right)^{2/3} T.
\end{equation}
The specific heats are calculated for a temperature of T=0.1 MeV, which is a
typical temperature for the inner crust matter at the cooling stage analysed here
(see the discussion below).
\begin{figure}[htb]
\includegraphics[scale=0.27,angle=-90]{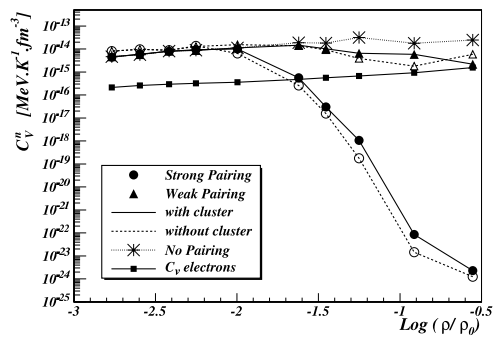}
\caption{Specific heat of the neutrons for the Wigner-Seitz cells listed 
in Table I. The results correspond to the strong and the weak pairing forces 
(see the text) and for the cells with (without) the nuclear clusters.
The specific heat of the non-uniform cells obtained when the pairing correlations
are switched off are indicated by star symbols. The square symbols show the 
specific heat of the electrons.}
\end{figure}
 From Figure 1 we can see that if the neutrons are in the normal phase, their
specific heat is greater than the specific heat of the electrons in all Wigner-Seitz 
cells.
When the neutron superfluidity is turned on, the specific heat of the neutrons 
is suppressed due to the pairing gap in the excitation spectrum. Since the 
suppression depends exponentially on the pairing gap, the results obtained with
the strong and the weak pairing forces are very different, as seen 
for the WS cells 1-5. For the second WS cell, in which the pairing gap in the
neutron gas region have the maximum value, the specific heat obtained with the
two pairing forces differs by about 7 orders of magnitude.
In the WS cells 7-10 the neutron gas is in the normal phase at the temperature T=100 
keV. Therefore both pairing forces give the same results for the specific heat.

In Figure 1 are shown also the values of the specific heat obtained when the nuclear
clusters are disregarded. For obtaining these values we have just removed the protons
from the cells and perform the FT-HFB calculations in the same conditions as for the
cluster+neutron gas. It can be seen that in some cases (see the results for the cells
2-4) the nuclear clusters could have a sizable influence upon the specific heat.
However, the influence of the nuclear clusters are relatively small compared to the
effect coming form the uncertainty of the pairing force.

\subsection{ Thermal diffusivity}

The specific heat enters in the heat transport through the 
thermal diffusivity, defined by $D=\frac{\kappa}{C_V}$, where $\kappa$ is the
thermal conductivity. In the inner crust, the latter is primarly determined
by the electrons. The dependence of thermal conductivity on density and temperature
was parametrized by Lattimer et al\cite{lattimer}, based on the calculations of 
Itoh et al \cite{itoh}. 
For a temperature above $10^8$ K analysed here, the conductivity is nearly independent
of the temperature and is given by $\kappa=C(\rho/\rho_0)^{2/3}$, where 
$C=10^{21}$ ergs cm$^{-1}$ s$^{-1}$. 
\begin{figure}[htb]
\includegraphics[scale=0.27,angle=-90]{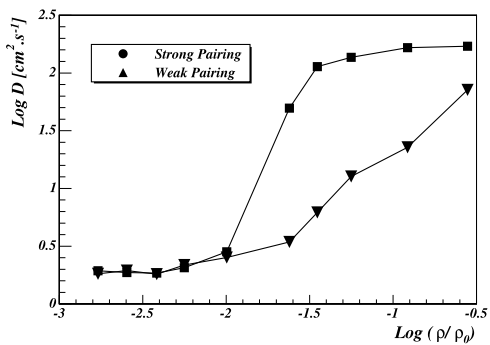}
\caption{Thermal diffusivity (neutrons plus electrons) corresponding to the
Wigner-Seitz cells listed in Table I. The notations are the same as in Figure 1.}
\end{figure}
With the conductivity given by this expression and
the specific heat calculated in the FT-HFB approach one gets the thermal diffusivity
shown in Figure 2.
As expected from the behaviour of the specific heat, the diffusivity
is much smaller for the weak pairing force, except  the last four
WS cells. For both pairing forces one can see that the diffusivity is much smaller
in the outermost layers of the inner crust. As seen below, these layers have an important
contribution to the cooling time of the inner crust.

\section{ Cooling time of  the inner crust matter}
   
 In order to calculate the cooling time, i.e., the time needed for the cooling wave to 
propagate from the cold core to the surface, one should integrate the heat equation
\begin{equation}
\frac{1}{r^2}\frac{\partial}{\partial r}[ r^2 \kappa \frac{\partial T}{\partial r}]
= C_V \frac{\partial T}{\partial t}.
\end{equation}
Since the specific heat and the conductivity depend on density and  
temperature profile of the crust, the solution of the heat equation is not trivial.
Here we use a simple model employed in Refs. \cite{brown,pizzochero}. The model is based
on the following assumptions: a) the spherical geometry for the heat transport is
approximated by a planar geometry, i.e, one considers the heat diffusion through a
one-dimensional piece of matter. This approximation is supported by the small thickness
of the inner crust compared to the size of the core; b) the inner crust is divided in
layers of constant thermal diffusivity.  The diffusion time through a layer of thickness
$x_i$ and diffusivity $D_i$ is calculated by the relation $t_i=\gamma \frac{x_i^2}{D_i}$ 
\cite{landau},
where the factor $\gamma$, which depends on the boundary conditions of the 
problem, is taken equal to $4/pi^2$ \cite{pizzochero}; c) the total diffusion time across
the crust is obtained by summing up the contributions of the layers, i.e.,
\begin{equation}
t_{diff}=\gamma \sum_i \frac{x^2_i}{D_i}.
\end{equation}

In the equation above the thermal diffusivity depends on density and temperature,
 $D_i=D(\rho(R_i),T(R_i))$, where $R_i$ is the position of the layer $i$.
 In the calculations we divide the inner crust into 10 layers, corresponding to the 
 10 cells listed in Table I. The position corresponding to each cell can be found by
 solving the Tolman-Oppenheimer-Volkov (TOV) equations, which provides the density
 profile of the star. In the present calculations we use the solution of TOV equations
 corresponding to the following equations of state \cite{vidana}: Baym-Pethick-Sutherland
 \cite{baym} for the outer crust, Negele-Vautherin \cite{negele} for the inner crust
 and Glendenning-Moszkowski \cite{glendening} for the core. From the solution of the
 TOV equations one extracts the radii $R_i$ corresponding to the densities of the
 cells given in Table I. Then, doing a linear interpolation, we determine the
 size $x_i$ of the layers considered for each cell. The results are shown in Table I.

 The diffusivity depends also on the temperature profile. Numerical simulations indicate
 that before the core-crust thermalisation the temperature is increasing from about T=0.1 MeV
 to about T=0.2-0.3 MeV when one goes from the outer part to the inner part of the crust.
 Since the inner part zones of the inner crust have large diffusivities, they contribute
 less to the cooling time compared to the outermost zones. Therefore, following Ref. \cite{pizzochero},
 we shall consider for all layers a flat temperature profile equal to T=0.1 MeV. 
 The diffusion time across the inner crust obtained for this value of the temperature
 is shown in Figure 3. The most striking thing we can notice is the critical dependence
 of the cooling time on the pairing force. Thus, for a strong pairing
 force the cooling time is about 12 years. The largest contributions come from the outermost
 zones, as noticed also in Ref. \cite{pizzochero}. Concerning the effect of the clusters, one can see that
 is rather small for this temperature. In the case of the weak pairing force, the cooling 
 time is increasing by about a factor two compared to the strong force. Moreover, if the neutron
 superfluidity is ignored completely, the cooling time is further increasing to about 90 years.
 These dramatic changes shows how important is the precise knowledge of the neutron matter superfluidity
 for the cooling time of the innner crust.

\begin{figure}[htb]
\includegraphics[scale=0.27,angle=-90]{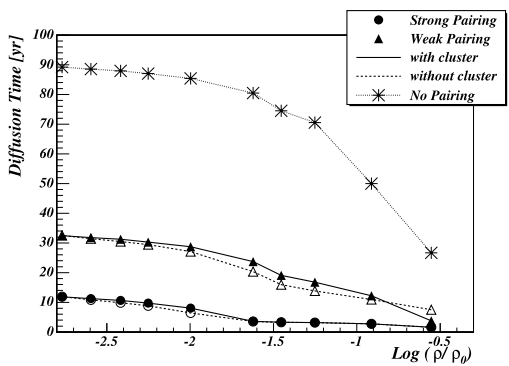}
\caption{The diffusion time across the inner crust. The notations
are the same as in Figure 1.}
\end{figure}

 A similar strong dependence of the cooling time on the pairing scenarios we have obtained by using two
 other zero range forces with the parameters fixed following a different protocol, i.e., a unique
 strength,  $V_0$=-648 MeV fm$^{-3}$, and two sets of parameters for the term dependend on density,
 $\eta$=\{0.95, 0.87 \} and $\alpha$=\{0.45, 0.2\}.  The value of $V_0$ was taken so that to get, for a
 smooth cut-off energy equal to 20 MeV, the experimental value for the scattering length of two free neutrons.
 The cooling times obtained with these two zero range pairing forces are equal to about 9.1 and 33.8 years,
 respectively.

 The cooling times calculated in this section are based on the assumption of  a flat temperature 
 across the inner crust. This is a rather drastic approximation, especially for the scenario of 
 a weak pairing force when, as seen in Figure 3, all regions of the inner crust contribute 
 significantly to the total diffusion time. More realistic calculations of the cooling time 
 should be based on dynamical solutions of the heat equations (9). 
\vspace*{0.2cm}
\section{Summary and Conclusions}

 We have estimated the cooling time of the inner crust matter using
specific heats calculated in the framework of HFB approach at finite temperature.
In order to study the effects of the neutron superfluidity on  thermal properties
of the inner crust, we have employed two paring forces. They have been fixed to
reproduce the pairing properties of infinite neutron matter given either by a
Gogny force or by microscopic calculations which take into acocunt polarisation effects.
For the latter we considered a maximum pairing gap in neutron matter equal to 1 MeV.
With the two pairing forces we have studied what are the effects of neutron
superfluidity on the specific heat and the heat difusion of inner crust matter.
It is shown that the heat difusion predicted by the two pairing forces are
rather different, especially in the higher density part of the inner crust.
These differences in the heat diffusion have a big influence upon the cooling time.
Thus, if one shifts from one pairing force to the other the cooling time is
changing by a factor of three. This show how large could be the window in which the
cooling time may vary due to the present lack of knowleadge of neutron matter
superfluidity.

The neutron superfluidity affects the cooling time through the specific heat,
calculated here with the non-collective quasiparticle spectrum provided by 
the FT-HFB equations. However, the excitation spectrum of the inner crust 
baryonic matter presents also low-lying collective modes \cite{supergiant}. 
Since these modes give an important contribution to the specific heat 
\cite{predeal}, they may also affect significantly the cooling time of 
the inner crust. 
This issue will be addressed in a future study.

{\bf Acknowledgment.} We thank Nguyen Van Giai and P. M. Pizzochero for valuable 
discussions during the completion of this work.

\end{document}